\documentclass[aps,pra,twocolumn,letterpaper,showpacs,floatfix,superscriptaddress]{revtex4}
\usepackage[dvips]{graphicx,color}
\usepackage{amsmath}

\newcommand{\half}{\frac{1}{2}}
\newcommand{\ket}[1]{\left|#1\right>} 
\newcommand{\bra}[1]{\left<#1\right|} 
\DeclareMathOperator{\tr}{Tr} 

\begin{document}

\title{Detecting antiferromagnetism of atoms in an optical lattice via optical Bragg scattering}

\author{T. A. Corcovilos}
\email{tac3@rice.edu}
\affiliation{Department of Physics and Astronomy, 
and Rice Quantum Institute,
Rice University, Houston, Texas 77005}
\author{S. K. Baur}
\affiliation{Laboratory of Atomic and Solid State Physics, Cornell University, Ithaca, New York 14853}
\author{J. M. Hitchcock}
\affiliation{Department of Physics and Astronomy,
and Rice Quantum Institute,
Rice University, Houston, Texas 77005}
\author{E. J. Mueller}
\affiliation{Laboratory of Atomic and Solid State Physics, Cornell University, Ithaca, New York 14853}
\author{R. G. Hulet}
\affiliation{Department of Physics and Astronomy,
and Rice Quantum Institute,
Rice University, Houston, Texas 77005}

\date{October 13, 2009}

\begin{abstract}

Antiferromagnetism of ultracold fermions in an optical lattice can be detected by Bragg diffraction of light, in analogy to the diffraction of neutrons from solid state materials.  A  finite sublattice magnetization will
lead to a Bragg peak from the $(\frac{1}{2}\,\frac{1}{2}\,\frac{1}{2})$ crystal plane with an intensity depending on details of the atomic states, the frequency and polarization of the probe beam, the direction and magnitude of the sublattice magnetization, and the finite optical density of the sample.  Accounting for these effects we make quantitative predictions about the scattering intensity and find that with experimentally feasible parameters the signal can be readily measured with a CCD camera or a photodiode and used to detect antiferromagnetic order.  
\end{abstract}

\pacs{37.10.Jk, 42.50.Ct, 71.10.Fd}

\maketitle

\section{Introduction}%
Ultracold atomic gases in optical lattices may be used as analog quantum simulators of 
 condensed matter models \cite{bloch08}.  Simulating the Fermi-Hubbard model is particularly interesting.  This model is nearly impossible to solve using traditional techniques, yet is amenable to simulation with cold atoms.  Most importantly, this model is relevant for understanding a wide class of strongly correlated electron systems: for example, many believe that it captures the essential physics responsible for high-temperature superconductivity in cuprates \cite{anderson02}.   The major technical challenges to this program are developing improved cooling and measurement techniques.  Here we explore how light scattering can be used to detect ``magnetic" order in a gas of fermionic atoms in an optical lattices, focussing on the antiferromagnetic order found at low temperatures in the insulating phase of the Hubbard model.  Producing (and detecting) such an antiferromagnetic state is a key step on the road to exploring superfluidity in the Hubbard model.

The Hubbard Hamiltonian for a two-spin-component Fermi gas is
$$
H = -t \sum_{\langle i,j\rangle, \sigma=\uparrow,\downarrow}
\left( c_{i,\sigma}^\dagger c_{j,\sigma} + \textrm{h.c.}\right)
+ U \sum_i n_{i,\uparrow}n_{i,\downarrow},
$$
where $i,j$ label lattice sites, angle brackets denote sums over nearest neighbors, $\sigma$ is the spin label, and the occupation of site $i$ is $n_{i,\sigma}=c_{i,\sigma}^\dagger c_{i,\sigma}$.  The coefficient $t$ is the hopping energy between adjacent sites, and $U$ is the on-site interaction energy.  Jaksch et al.~\cite{jaksch98} show how a gas of atoms in an optical lattice reduces to this model at low temparatures, and provided expressions for $t$ and $U$ in terms of atomic properties.
We consider the case $U,t>0$.

At half-filling (one atom per lattice site) and strong repulsion ($U \gg t $), the Hamiltonian reduces to an antiferromagnetic Heisenberg model:
\begin{equation}\label{eq:Hafm}
H_\textrm{AFM} = J \sum_{\langle i,j \rangle} \mathbf{S}_i \cdot \mathbf{S}_j,
\end{equation}
with $J\equiv4t^2/U$ and $\mathbf{S}_i$ is the spin operator for site $i$.  At lower lattice depths, $J$ is renormalized to a slightly lower value by the direct interaction between atoms on neighboring sites \cite{mathy09}. An infinite system described by $H_\textrm{AFM}$ undergoes a second order phase transition to an antiferromagnetically ordered N\'eel state at a temperature $T_N \sim J \ll U$.  

Achieving these temperatures is far from trivial.
Experiments on cold fermions \cite{jordens08,schneider08} have reached sufficiently low temperatures to observe
characteristics of the Mott insulator, but so far $T>T_N$.  
Much lower temperatures will be reached in the near future, when the next generation of cooling protocols are implemented \cite{kent05,koetsier08,ho08,j.ho09}.  Many of these approaches are interesting not only for their utility, but also for the insight they provide about fundamental issues of thermodynamics and quantum statistics.

The proposals for detecting antiferromagnetic ordering in an atomic gas  mostly rely on the fact that appropriately tuned light couples differently to the two spin components \cite{partridge06, shin06}.  For example, in an experiment where one can directly image individual lattice sites  \cite{nelson07,gillen09}, one can use this selectivity to directly resolve the antiferromagnetic ordering.  An alternative proposal involves measuring the spatial noise correlations in images of the density profile after turning off the trapping and lattice potentials \cite{altman04,polkovnikov06,bruun09}.  The experimental noise contains information about the density-density correlation functions of the system, and in principle encodes the antiferromagnetism. This method has been used to detect imposed lattice order in a bosonic Mott insulator \cite{foelling05,  spielman07} and a fermionic band insulator \cite{rom06}. Its disadvantage is sensitivity to technical noise and that averaging over many experimental shots under similar  conditions  is  needed to  obtain  sufficient statistics.

Here we propose detecting magnetic order through \emph{in-situ} Bragg diffraction of \emph{light} (as opposed to Bragg diffraction of \emph{atoms}, another common technique \cite{kozuma99}). This method is analogous to  neutron scattering \cite{shull,squires} and magnetic x-ray scattering \cite{platzman70,bruneland81}, as used in solid state physics.   Optical Bragg diffraction  has been used previously in cold atomic gases to confirm the crystalline ordering of nondegenerate atoms in optical lattices  \cite{birkl95, weidemueller95, weidemueller98}.
Compared to the other proposed techniques, the primary advantages of Bragg diffraction are its simplicity, speed, and relatively large signal.  In this work, we report a detailed theoretical analysis of Bragg scattering from an array of atoms trapped in an optical lattice and demonstrate its usefulness under typical experimental conditions.

\section{Basic Theory}
We consider atoms confined in a three dimensional simple cubic optical lattice potential,
$
V(x,y,z) = V_0 [\sin^2(\pi x/a) + \sin^2(\pi y/a) + \sin^2(\pi z/a)],
$
where the lattice constant $a$ is half of the lattice laser wavelength and $V_0$ is the depth of the optical potential, assumed to be large enough that the tight-binding approximation holds.  Although the transverse Gaussian intensity profile of the laser beams causes $V_0$ to be position dependent, in the relevant part of the cloud it can be taken to be constant.

Two collisionally closed atomic hyperfine sublevels ($\uparrow$, $\downarrow$) may be treated as a pseudospin-$\frac{1}{2}$ system, where there is no mechanism for spin relaxation.  We assume that  the two states are separated in energy by a splitting $2\Delta_0$, and have optical resonance transitions suitable for imaging with line width $\Gamma$ -- concrete examples using $^6$Li will be given later.  An external magnetic field is applied to tune the interactions between atoms via a Feshbach resonance.  This field also defines the quantization axis for the Zeeman sublevels of the hyperfine states and, hence, of the pseudo-spin states.
\subsection{Bragg scattering cross section in the Born approximation}
\begin{figure}
	\centering
		\includegraphics{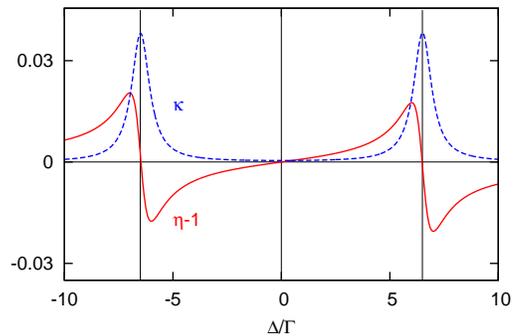}
	\caption{\label{fig:ref}(Color online) Real ($\eta-1$, solid red line) and imaginary ($\kappa$, dashed blue line) parts of the index of refraction for a uniform gas of $^6$Li atoms as a function of laser detuning.  The atoms are taken to be in the two lowest hyperfine states with densities $n_\downarrow=n_\uparrow=1/[2({532}\,\mathrm{nm})^3]=3.3\times 10^{12}\,\mathrm{cm^{-3}}$ in an external magnetic field of ${834}\,\mathrm{G}$, yielding a Zeeman splitting of $2\Delta_0=2\pi\times{76}\,\mathrm{MHz}$.  The laser excites the $\mathrm{D_2}$ line with linewidth $\Gamma=2\pi\times{5.9}\,\mathrm{MHz}\approx 2 \Delta_0/13$.	Thin vertical lines denote the two resonances and their midpoint.} 
	\end{figure}

We initially consider Bragg scattering in the limit of low probe intensity and low optical density.  In this limit the Born approximation may be used.  The more general case will be considered in Sec.~\ref{sec:multscatt} and the appendix.  Our system consists of atoms in an optical lattice described by operators for position $\hat{\mathbf{r}}_j$ and occupation number $\hat{n}_{j\sigma}$. In the Born approximation, the total amplitude for elastic scattering reduces to \cite{loudon}
\begin{equation}
\label{eq:bornscatt}
{F}_{\mathbf{k}_f,\mathbf{k}_i}=\sum_j \sum_{\sigma=\uparrow,\downarrow} f_{\sigma} \hat{n}_{j\sigma} e^{i \mathbf{K}\cdot \hat{\mathbf{r}}_j},
\end{equation}
where the difference between incoming and outgoing wavevectors of the photons is $\mathbf{k}_i-\mathbf{k}_f=\mathbf{K}$, and
\begin{align}
f_{\sigma}&=-\frac{3}{2 k} (\mathbf{e}^{*}_{\mathbf{k}_f,\lambda_f} \cdot \mathbf{e}_m)
(\mathbf{e}_m^{*} \cdot \mathbf{e}_{\mathbf{k}_i,\lambda_i}) \frac{\Gamma/2}{\Delta_{\sigma}+i \Gamma/2} \notag \\
&=\frac{3 }{2 k} (\mathbf{e}^{*}_{\mathbf{k}_f,\lambda_f} \cdot \mathbf{e}_m)
(\mathbf{e}_m^{*} \cdot \mathbf{e}_{\mathbf{k}_i,\lambda_i}) {e^{i \delta_{\sigma}} \sin(\delta_{\sigma})},\label{eq:fj}
\end{align}
is the scattering amplitude for individual spin-$\sigma$ atoms. Here, $\mathbf{e}_{\mathbf{k}_i,\lambda_i}$ is the polarization vector of the incoming photon with wavevector $k_i$ and polarization $\lambda_i$ (generally elliptical), and $\mathbf{e}_m$ is the polarization which couples to the resonant optical transition.  The phase shifts $\delta_{\sigma}$ are related to the linewidth and detuning by $\tan(\delta_\sigma) = -\Gamma/2\Delta_\sigma$.  
Note that Eq.~\eqref{eq:bornscatt} is only valid when the optical density of the sample is small compared to one. Optically dense samples will be considered in Sec.~\ref{sec:multscatt}.  To achieve spin dependent scattering the probe light frequency must be near an atomic resonance.  Hence the interaction between light and the atoms is neither purely diffractive nor purely absorbtive.  This is clearly illustrated by considering the lattice free case and assuming the atoms are uniformly distributed with densities $n_\uparrow$ and $n_\downarrow$.  The electric susceptibility $\chi_\textrm{bulk}$ of the bulk sample is then \cite{loudon},
$$
\chi_{\textrm{bulk}} = -\frac{4\pi c^3}{\omega^3} \left(\frac{f_{\uparrow} n_\uparrow + f_{\downarrow} n_\downarrow}{2}\right),
$$
where $\omega$ is the angular frequency of the light, $c$ is the speed of light in vacuum.  The real and imaginary components of the index of refraction $\eta$ and $\kappa$ are then given by the usual relationship: $\eta+i\kappa=\sqrt{1+\chi}$.  The dependence of these quantities on the probe laser frequency for $^6$Li is show in Fig.~\ref{fig:ref} in the  case where $n_\uparrow=n_\downarrow$.

As seen in the figure, when the probe detuning is halfway between the levels, $\eta$ is unity and $\kappa$ has a local minimum. For this detuning the scattering phase shifts of the two states obey $\delta_{\uparrow}=-\delta_{\downarrow}$.  This will be the operating point of our proposed measurement.
There are three advantages to working at this point:
(1) we can neglect bulk dispersive effects because $\eta=1$, (2) absorption is small, and (3) we are maximally sensitive to the spin order.  

Experiments measure the differential scattering cross-section 
\begin{eqnarray*}
\frac{d\sigma(\mathbf{K})}{d\Omega}&=&\tr\left[ \hat{\rho}_i \hat{F}^{\dagger}_{\mathbf{k}_f,\mathbf{k}_i} \hat{F}_{\mathbf{k}_f,\mathbf{k}_i}\right]\\
&=& \frac{9}{4 k^2} \sum_{\lambda_f}\left|(\mathbf{e}^{*}_{\mathbf{k}_f,\lambda_f} \cdot \mathbf{e}_m) (\mathbf{e}^{*}_m \cdot \mathbf{e}_{\mathbf{k}_i,\lambda_i}\!)\right|^2 \\
& & \times \sum_{\sigma,\sigma',j,j'} \Big[ \left\langle \hat{n}_{j\sigma} \hat{n}_{j'\sigma'} e^{i \mathbf{K}(\hat{\mathbf{r}}_j-\hat{\mathbf{r}}_{j'})} \right\rangle \\
& & \times e^{i (\delta_{\sigma}-\delta_{\sigma'})} \sin(\delta_{\sigma}) \sin(\delta_{\sigma'})\Big].
\end{eqnarray*}
The density matrix $\hat{\rho}_i=\hat{\rho}_A \otimes  \ket{\mathbf{k}_i,\lambda_i} \bra{\mathbf{k}_i,\lambda_i}$ describes the initial state of the system (where $\hat{\rho}_A$ is the thermal density matrix of the atoms). Tracing over outgoing polarizations gives a factor $\sum_{\lambda_f}|(\mathbf{e}^{*}_{\mathbf{k}_f,\lambda_f} \cdot \mathbf{e}_m)
(\mathbf{e}_m^{*} \cdot \mathbf{e}_{\mathbf{k}_i,\lambda_i})|^2=\frac{1}{4}(1+\cos^2\theta)(1+\cos^2\theta')$, where $\theta$ $(\theta')$ is the angle between the magnetization axis and the incoming (outgoing) wavevector, assuming the incoming elliptical polarization is chosen to maximize the signal
\footnote{The optimal incoming polarization is found by projecting the transition polarization $\mathbf{e}_m$ onto the 2d space $\perp\mathbf{k}$.  Defining the unit vectors $\hat{s} \propto \hat{k}\times\hat{z}$ and $\hat{p} \propto \hat{k}\times\hat{s}$, the optimal incoming polarization is $\mathbf{e}_{k,\lambda}=(\hat{s}-i\cos\theta \;\hat{p})/\sqrt{1+\cos^2\theta}$.}. 
The thermal average $\left \langle \hat{n}_{j\sigma} \hat{n}_{j'\sigma'} e^{i \mathbf{K}(\hat{\mathbf{r}}_j-\hat{\mathbf{r}}_{j'})} \right\rangle$ factorizes with $\left \langle \hat{n}_{j\sigma} \hat{n}_{j'\sigma'} \right \rangle=\left \langle\left(\frac{1}{2}+\sigma \hat{S}_{zj} \right) \left(\frac{1}{2}+\sigma' \hat{S}_{zj'}\right) \right \rangle$, where we assume half-filling $n_{i\uparrow}+n_{i\downarrow}=1$, $\sigma=\pm 1$, and $S_{zi}=\frac{1}{2}(n_{i\uparrow}-n_{i\downarrow})$). The factor
$\left \langle e^{i \mathbf{K}(\hat{\mathbf{r}}_j-\hat{\mathbf{r}}_{j'})}\right \rangle \approx e^{-2 W} e^{i \mathbf{K}(\mathbf{R}_j-\mathbf{R}_{j'})}$, where $e^{-2 W}=e^{-l^2 K^2/2}$ is the Debye-Waller factor due to the zero-point motion of the atoms around the lattice sites $\mathbf{R}_j$ in terms of the harmonic oscillator length $l\approx \sqrt{\hbar/m\omega}$ of the individual wells. The differential cross-section becomes
\begin{align}
\frac{d\sigma(\mathbf{K})}{d\Omega}
=&  \frac{9}{4 k^2} \frac{1}{4}(1+\cos^2\theta)(1+\cos^2\theta') e^{-2 W}\notag \\
&\times  \Big( \alpha(\delta_{\uparrow},\delta_{\downarrow}) C(\mathbf{K}) 
  + \beta(\delta_{\uparrow},\delta_{\downarrow}) S(\mathbf{K}) \Big), \label{eq:cs}
\end{align}
where we introduce the crystal structure factor \mbox{$C(\mathbf{K})=\sum_{i,j}e^{i\mathbf{K}\cdot(\mathbf{R}_i-\mathbf{R}_j)}$} and spin structure factor $S(\mathbf{K})=\sum_{i,j} e^{i\mathbf{K}\cdot(\mathbf{R_i}-\mathbf{R_j})} \langle S_{zi} S_{zj} \rangle$.  The coefficients are
$
\alpha(\delta_{\uparrow},\delta_{\downarrow})=\frac{1}{4}|\bar{f}_{\uparrow}+\bar{f}_{\downarrow}|^2$ and
$\beta(\delta_{\uparrow},\delta_{\downarrow})=|\bar{f}_{\uparrow}-\bar{f}_{\downarrow}|^2
$
with $\bar{f}_{\sigma} = e^{i \delta_{\sigma}} \sin(\delta_{\sigma})$. As described previously, for a laser frequency where $\delta_{\uparrow}=-\delta_{\downarrow}\equiv\delta \ll 1$ 
(this is achieved when $\Delta_{\uparrow}=-\Delta_{\downarrow}=\Delta_0$), 
the coefficients simplify:
\begin{eqnarray}
\alpha(\delta_{\uparrow},\delta_{\downarrow})&=&\sin^4(\delta)\approx \delta^4 \approx (\Gamma/2\Delta_0)^4,\notag \\
\beta(\delta_{\uparrow},\delta_{\downarrow})&=&\sin^2(2 \delta)\approx 4 \delta^2 \approx 4(\Gamma/2\Delta_0)^2.\notag
\end{eqnarray}

The factor $S(\mathbf{K})$ in Eq.~\eqref{eq:cs} represents the well-known result that (spin-selective) Bragg scattering  measures the spin structure factor. In solids one uses neutron scattering for the same purpose and the coupling between the magnetic moment of the neutrons and electron spin causes different scattering amplitudes for $\uparrow,\downarrow$ electrons \cite{ashcroft}. In cold atoms the situation is even more favorable as this result shows: by choosing detuning with opposite phase shifts for the two spin states one can basically turn off the signal from the underlying lattice and probe only the spin order (i.e.~$\alpha\ll\beta$).

The predicted locations of Bragg diffraction peaks (i.e.~maxima in ${d\sigma(\mathbf{K})}/{d\Omega}$) are given by the Laue condition: the difference in wavevector between the outgoing and incoming scattered light ($\mathbf{k}_i-\mathbf{k}_f=\mathbf{K}$) is equal to a reciprocal lattice vector $\mathbf{Q}$ of the crystal  \cite{als-nielsen}.  Rewriting this condition in terms of incoming and outgoing unit propagation vectors $\mathbf{\hat{k}}_i, \mathbf{\hat{k}}_f$, resp. and the wavelength of the scattered light $\lambda$:
\begin{equation}\label{eq:laue}
\mathbf{\hat{k}}_f-\mathbf{\hat{k}}_i =
\frac{\lambda}{a} \left(l\;m\;n\right),
\end{equation}
where the Miller indices of the scattering plane are defined for a cubic lattice by $\left(l\,m\,n\right) \equiv \mathbf{Q} a/2\pi$.

Additionally, the magnitude of the reciprocal lattice vectors available for Bragg scattering is further restricted by the  the vector triangle inequality:
\begin{gather}
\big| |\mathbf{k}_f|-|\mathbf{k}_i| \big|
\leq |\mathbf{Q}| \leq
|\mathbf{k}_f|+|\mathbf{k}_i|,\notag \\
0\leq|\mathbf{Q}|\leq2k,\label{eq:triangle}
\end{gather}
where the second line assumes elastic scattering.  The consequence of the above expression is that for the configurations typical of optical lattice experiments  where $a\sim\lambda$ only a handful of choices for $\mathbf{K}$ will produce Bragg scattering.

\subsection{Finite simple cubic lattice}\label{sec:cubic}
The structure factors $S(\mathbf{K})$ and $C(\mathbf{K})$ in Eq.~\eqref{eq:cs} are sums that may be performed exactly for a simple cubic lattice with length per side of $L$ sites ($N=L^3$ total sites).  By recognizing that the sum in the crystal structure factor is that of a geometric sequence in this configuration, it evaluates to
\begin{equation}\label{eq:CK}
C(\mathbf{K})=\prod_{j=x,y,z}
\frac{\sin^2(K_j a L/2)}{\sin^2(K_j a /2)}.
\end{equation}
If $\mathbf{K}$ corresponds to a reciprocal lattice vector of the crystal then the sum is maximized with $C(\mathbf{K})=N^2$, demonstrating the coherent enhancement of the scattering when the Laue condition is satisfied.  As $\mathbf{K}$ moves away from a reciprocal lattice vector, $C(\mathbf{K})$ falls off approximately quadratically. 

The spin structure factor $S(\mathbf{K})$ may be evaluated in a similar way.  Magnetically ordered phases may be described by a spin ordering vector $\mathbf{q}$ such that the average staggered spin is $\mathbf{s} = \frac{1}{N} \sum_j \mathbf{S} e^{i\mathbf{q}\cdot\mathbf{R}_j}$, having $z$-component $s_z$.  In the mean field approximation the spin dependent factor in $S(\mathbf{K})$ is
$\langle S_{zi}S_{zj} \rangle 
= \langle S_{zi}\rangle\langle S_{zj} \rangle
= s_z^2 e^{i\mathbf{q}\cdot(\mathbf{R}_i-\mathbf{R}_j)}$. 
In particular, a ferromagnetic (fully polarized) phase has $\mathbf{q}=0$ and a N\'eel AFM has $\mathbf{q}=\frac{2\pi}{a}(\half\;\half\;\half)$.  With this notation
\begin{equation}\label{eq:SK}
S(\mathbf{K})=s_z^2\prod_{j=x,y,z}
\frac{\sin^2[(K_j+q_j) a L/2]}{\sin^2[(K_j+q_j) a /2]}.
\end{equation}
When $\mathbf{K}+\mathbf{q}$ corresponds to a reciprocal lattice vector, $S(\mathbf{K})$ is maximized with value $s_z^2N^2$. 

The locations of the maxima in $C(\mathbf{K})$ and $S(\mathbf{K})$ are well-known results from solid state physics.  
As explained in standard textbooks \cite{ashcroft,als-nielsen}, Bragg scattering for a simple cubic lattice occurs only at planes where the Miller indices are integers in the absence of magnetic ordering.
If the crystal contains antiferromagnetic ordering, half-integer Miller indices are also possible, indicating a doubling of the crystal unit cell.
Planes with mixtures of integer and half-integer Miller indices have zero scattering amplitude.

In the case of a paramagnetic phase (e.g.~when $T>T_N$), the spins $S_{iz}$ appearing in the sum $S(\mathbf{K})$ do not show long range order.   
We can then estimate the diffuse background by assuming that all the spins point in random directions:
$S(\mathbf{K})=\left<\sum_i S_{iz}e^{i\mathbf{K}\cdot\mathbf{R}_i}\right>^2\approx N/4$, independent of $\mathbf{K}$, i.e.~this term describes incoherent scattering.  
For a real experiment at finite temperature, a portion of the sample is likely to be in the paramagnetic phase.  The incoherent scattering from these disordered spins will yield a diffuse background.

\section{Example: $^6\textrm{Li}$ in a red-detuned lattice}
To demonstrate this technique for a typical experimental configuration, we simulate a crystal of $^6\textrm{Li}$ atoms evaporatively cooled to Fermi degeneracy in a far off-resonance optical trap, then loaded into a 3d simple-cubic lattice created from three orthogonal far red detuned laser plane-wave standing waves at ${1064}\,\mathrm{nm}$.  This experiment is currently underway at Rice University.

The $^6\mathrm{Li}$ atoms are prepared in an incoherent balanced mixture of the $F=\half$, $m_F=\pm \half$ Zeeman sublevels of the $2S_{\half}$ atomic ground state (denoted $\downarrow$ and $\uparrow$, respectively).  These two sublevels play a role analogous to the spin-$\half$ electrons in crystalline solid.  An externally applied uniform magnetic field controls interatomic interactions via a Feshbach resonance \cite{houbiers98}.  We choose the direction of the magnetic field to be along one of the lattice axes for experimental convenience.  This field also determines the quantization axis of the atomic states.

\begin{figure}
\centering
\includegraphics[scale=0.5]{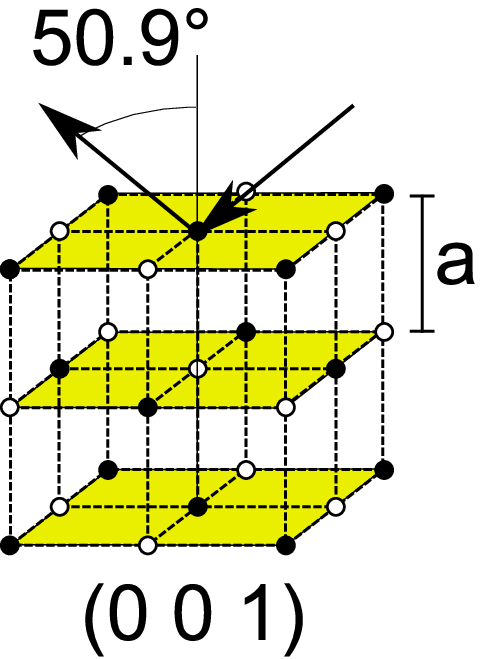}%
\includegraphics[scale=0.5]{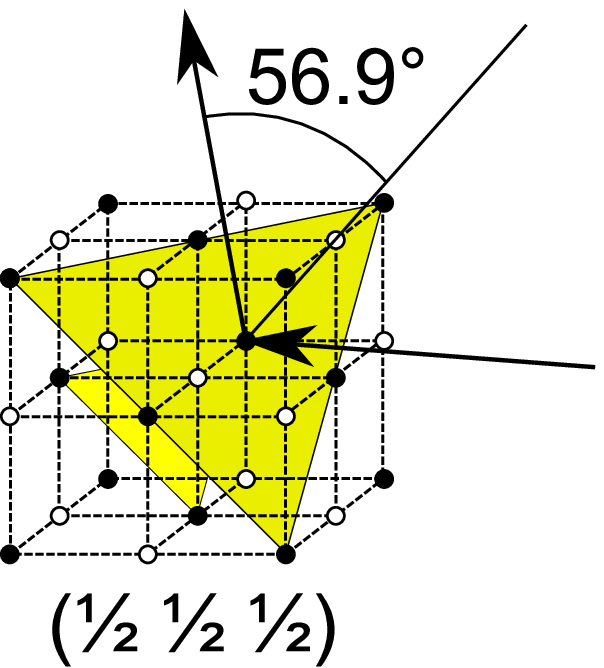}%
\caption{\label{fig:cube}(Color online) Geometry for scattering from $(0\;0\;1)$ plane (left) and $(\half\;\half\;\half)$ plane (right) for a cubic lattice with lattice spacing $a$.  The scattering planes are shaded.  The thin solid line is normal to the scattering plane.  The color of the lattice sites indicates spin; this figure shows AFM ordering.  The arrows represent the directions of the incoming and outgoing scattered light.  The labeled angles are between the outgoing light and the normal, which is the same value as the angle between the incoming light and the normal.  The values of the angles are given by the Bragg condition: $\cos^{-1}(|\mathbf{Q}|/2|\mathbf{k}|)$, assuming lattice spacing of $a={532}\,\mathrm{nm}$ and probe wavelength of ${671}\,\mathrm{nm}$.}
\end{figure}

The probe light is tuned near the $2S_{\half}\rightarrow2P_{\frac{3}{2}}$ resonance of $\mathrm{^6Li}$ with wavelength $\lambda = {671}\,\mathrm{nm}$.  Because $\lambda$ is similar in magnitude to the lattice constant $a$, only a few lattice planes will produce diffraction.  Applying Eq.~\eqref{eq:triangle}:
\begin{equation*}
0\leq l^2+m^2+n^2 \leq \left(\frac{2a}{\lambda}\right)^2 \approx 2.51.
\end{equation*}
The only valid nonmagnetic scattering planes for our system are $({\pm}1,0,0)$, $({\pm}1,{\pm}1,0)$ and permutations of these indices. (The $(0\,0\,0)$ case represents unscattered light).
The allowed magnetic scattering planes are $({\pm}\half,{\pm}\half,{\pm}\half)$.    The basic geometry for two of the scattering planes is depicted in Fig.~\ref{fig:cube}, with the values of the scattering angles shown.
In an experiment one could probe a nonmagnetic peak such as $(0\;0\;1)$ to confirm the cubic ordering of the sample and a magnetic peak such as $(\half\;\half\;\half)$ to detect the AFM ordering.  The ratio of these two peak intensities is a measure of the staggered magnetization, proportional to $s_z^2$ and independent of $N$.

The Zeeman splitting between states $\uparrow$ and $\downarrow$ is $2\Delta_0=2\pi\times{76}\,\mathrm{MHz} \approx 13\Gamma$ when the external magnetic field is tuned near the Feshbach resonance at ${834}\,\mathrm{G}$.  The bulk optical properties of the sample are those shown earlier in Fig.~\ref{fig:ref} for one atom per site and equal spin populations.

To estimate the optical signal strength produced by the scattering we consider a detector (CCD array or photodiode) placed $r={0.5}\,\mathrm{m}$ away from the atoms along the direction of the outgoing wavevector where we expect a peak.  We then numerically evaluate Eq.~\eqref{eq:cs}, assuming a perfectly ordered sample of  $N=41^3$ atoms and a lattice depth of $V_0 = {10}\,\mathrm{\mu K}$ (about 7 lattice photon recoil energies). The value of the spin structure factor varies depending on the phase we wish to calculate.  For a fully polarized phase the spin ordering vector is chosen to be $\mathbf{q}=0$.  For the AFM phase, we use $\mathbf{q}=\frac{2\pi}{a}(\half\,\half\,\half)$.  Finally, for the paramagnetic state we assume the spin structure factor is independent of direction with a value of $S(\mathbf{K})=N/4$, as mentioned previously in Sec.~\ref{sec:cubic}.  The maximum light intensity for the Bragg peak is given by
$I_\textrm{out}(\mathbf{k}',r)=\frac{I_\textrm{in}}{r^2} \frac{d\sigma(\mathbf{k}')}{d\Omega}$, where $\frac{d\sigma}{d\Omega}$ denotes the differential cross-section  in direction $\mathbf{k}^\prime$.  We take the  probe laser to be a plane wave with intensity $I_\textrm{in}={0.5}\,\mathrm{mW\,cm^{-2}}$ and beam waist much larger than the sample size. The total scattered power is estimated by numerically integrating over the area of the detector from the center of the peak radially to the first minimum.  The results of the calculations for the various phases and the two scattering planes of interested are given in
Table \ref{tab:pred}. For the maximal scattering configurations we find approximately ${1}\,\mathrm{nW}$ of power in the scattered beam spread over about ${10}\,\mathrm{mrad}$ of angular divergence.  Using commerically available photodiodes (with an appropriate collecting lens) and integration times of ${10}\,\mathrm{\mu s}$ should be sufficient to detect this power.  For higher sensitivity, inexpensive CCD cameras may be used \footnote{For example, the Basler Scout series, model SCA1000-20fm CCD camera has a sensitivity of about 12 photons per count at ${671}\,\mathrm{nm}$ and a noise floor of about 80 photons at this wavelength.}.
\begin{table}
	\centering
	\caption{\label{tab:pred}Predicted experimental values of maximum scattering cross-section, peak intensity, and scattered power for various magnetic states (paramagnet [PM], antiferromagnet [AFM] and polarized [Pol.]) and scattering planes using typical parameters, calculated as described in the text}
	\begin{tabular}{c|c c c}
	 & PM & AFM & Pol. \\
	 \hline
	 & \multicolumn{3}{c}{$(0\;0\;1)$ scattering plane} \\
	 $\frac{d\sigma}{d\Omega}\;(\mathrm{cm^2})$ & $1.4\times10^{-5}$ & $1.4\times10^{-5}$ & $2.3\times10^{-3} $ \\
	 I ($\mathrm{W/cm^2}$) & $2.7\times10^{-12}$ & $2.7\times10^{-12}$ & $4.5\times10^{-10}$ \\
	 P (W) & $5.7\times10^{-12}$ & $5.7\times10^{-12}$ & $9.4\times10^{-10}$ \\
	 & \multicolumn{3}{c}{$(\half\;\half\;\half)$ scattering plane} \\
	 $\frac{d\sigma}{d\Omega}\;(\mathrm{cm^2})$ & $4.7\times10^{-8}$ & $3.2\times10^{-3} $ & $7.3\times10^{-13}$ \\
	 I ($\mathrm{W/cm^2}$) & $9.4\times10^{-15}$ & $6.5\times10^{-10}$ & $1.5\times10^{-19}$ \\
	 P (W) & $2.0\times10^{-14}$ & $1.4\times10^{-9}$ & $9.3\times10^{-19}$ \\
	\end{tabular}
\end{table}

By varying the direction of the incoming beam, we find that  samples of this size are sensitive to the direction of the incoming beam on the $\mathord\sim{10}\,\mathrm{mrad}$ scale.  Given the size of our simulated crystal, this is consistent with the expected diffraction limit.

\subsection{Finite temperature}\label{sec:temp}
\begin{figure}
	\centering	\includegraphics{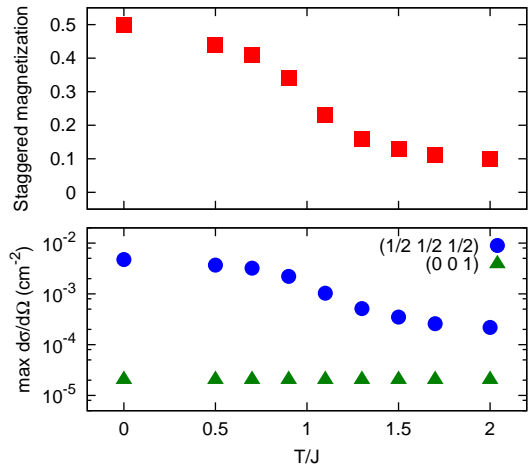}%
	\caption{\label{fig:pint2}(Color online) Staggered spin per atom of the system (red squares) calculated using a quantum Monte Carlo simulation of $6^3$ atoms, and maximum scattering cross-sections of the magnetic and nonmagnetic Bragg peaks (blue circles and green triangles, resp.) as a function of temperature.  The cross-sections are calculated using Eq.~\eqref{eq:cs} with the above value of staggered magnetization for a system of $41^3$ atoms.}
\end{figure}

The previous analysis assumed that the sample was completely ordered.  At finite temperature, espectially for $T$ near $T_N$, the AFM ordering will be incomplete.    To the estimate effect of finite temperature on the Bragg scattering intensity, we evaluate Eqs.~\eqref{eq:cs}, \eqref{eq:CK}, and \eqref{eq:SK} for a system with $L=41$, assuming the staggered magnetization as calculated by a quantum Monte-Carlo calculation for a smaller system with $L=6$.  While providing an estimate of the role of thermal fluctuations, our procedure of scaling up the 216 site result to 68921 sites will overestimate the finite-size smearing of the phase boundary.  We used the ALPS (Algorithms and Libraries for Physical Simulations) package to calculate this magnetization \cite{albuquerque07}.  As shown in Fig.~\ref{fig:pint2}
this extrapolation predicts that
the $(\half\,\half\,\half)$ Bragg peak begins to appear as the temperature goes below $T_N=0.95 J$, which is consistent with the bulk behavior of the system~\cite{sandvik98}.  By measuring the ratio of the two Bragg peak intensities we gain a measurement of $s_z$ independent of shot-to-shot fluctuations in $N$.
	
\section{Further considerations}
\subsection{Multiple Scattering}\label{sec:multscatt}
For optically dense samples multiple scattering cannot be neglected when calculating the differential cross-section.  We calculate the effect of multiple scattering by numerically solving the Lippmann-Schwinger equation, which is derived in the appendix. The key result is the scattering amplitude (also Eq.~\eqref{eq:fullF})
\begin{eqnarray}
F_{\mathbf{k_f},\lambda_f;\mathbf{k_i},\lambda_i}
&=&-\frac{3}{2 k}\frac{\hbar \Gamma}{2} \left( \mathbf{e}_{\mathbf{k}_f,\lambda_f} \cdot \mathbf{e}_m^{*}\right) \left( \mathbf{e}_m \cdot \mathbf{e}_{\mathbf{k}_i,\lambda_i}^{*}\right)\notag \\
& &\times \sum_j e^{-i \mathbf{k}_f \cdot \mathbf{r}_j}  A_j,\label{eq:fullF2}
\end{eqnarray}
where $A_j$ are numerical coefficients found by solving Eq.~\eqref{eq:lippmannschwinger}.

Using these results, we qualitatively study the consequences of finite optical density (OD), in particular the interplay with the position uncertainty of the atoms. (The Debye-Waller factor in Eq.~\eqref{eq:cs} is an approximation of this effect \cite{ashcroft}.)  In this section we consider a spherical sample and assume that all lattice sites within a radius $R$ are occupied. Position uncertainty caused by zero-point motion of the atoms is modeled as random displacements of the (point-like) atoms by a Gaussian distribution with standard deviation $\sigma=\Delta x \times a$. For the ground state of a sinusoidal well in the tight-binding regime $\Delta x=1/\pi V_0^{1/4}$, where $V_0$ is the lattice depth in units of the lattice recoil energy.  For a typical experimental value of $V_0=7$, $\Delta x = 0.2$.  The lattice spacing is $a=532\,\mathrm{nm}$ and the wavelength of the probe beam is $\lambda=671\,\mathrm{nm}$.  
\begin{figure}
	\centering
	\includegraphics{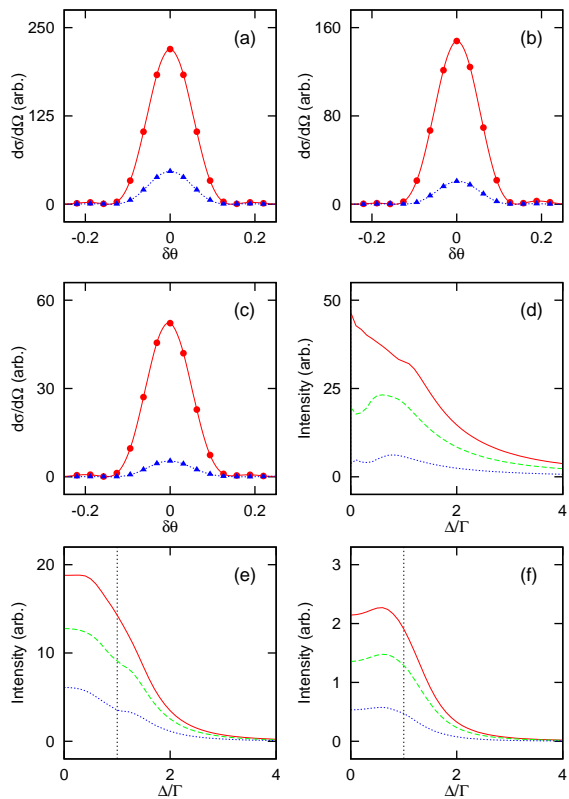}
	\caption{(Color online)\label{fig:multscatt}(a--c): Differential cross-section of the $(0\,0\,1)$ Bragg peak for a fully polarized spherical sample with $N=925$ atoms ($\mathrm{OD}=37$), resonant probe light, and varying mean displacements (disorder) $\Delta x=\sigma/a=0, \;0.1, \;0.2$ (corresponding to (a), (b), (c)) calculated from Eqs.~\eqref{eq:fullF2} and \eqref{eq:lippmannschwinger} (blue triangles), as a function of outgoing direction $\delta\theta$ relative to the peak center (directions shown in Fig.~\ref{fig:cube}). For comparison we also show the Born approximation result (red circles).  The lines are guides to the eye.
(d): Intensity of the $(0\;0\;1)$ Bragg peak as a function of detuning $\Delta$ in the fully polarized case  $\Delta x=0, \;0.1, \;0.2$ (solid red, dashed green, dotted blue lines).  The resonance is at $\Delta=0$.
(e--f):  Intensity of the $(\half\;\half\;\half)$ Bragg peak as a function of $\Delta$ for a N\'eel AFM state with Zeeman splitting between spin states of $2\Delta_0=2\Gamma$ for varying  $\Delta x=0, \;0.1, \;0.2$ (solid red, dashed green, dotted blue lines) and particle numbers (e): $N=925$ ($\mathrm{OD}=7.4$ at $\Delta=0$) and (f): $N=257$ ($\mathrm{OD}=4.8$ at $\Delta=0$).
The data in (d--f) have been symmetrized about $\Delta = 0$ to remove the effect of residual polarization caused by the small sample size.
The vertical dotted lines in (e, f) indicate the atomic optical resonance.}
\end{figure}

{\em Fully polarized:}
We first consider a model for a fully polarized state (all atoms in the same pseudospin state) with $R=6 a$ ($N=925$ atoms).  The angular distribution of the differential cross-section is shown in Fig.~\ref{fig:multscatt}(a--c) for various $\Delta x$ using both multiple-scattering and Born approximation methods.  The width of the peak is consistent with the diffraction limit for a sample of this size.  As shown in Fig.~\ref{fig:multscatt}(d), we find that for high optical density, the intensity of the $(0\; 0\; 1)$ Bragg peak develops a local minimum at resonance (zero detuning).  On resonance the optical density is highest, and even though the optical scattering cross-section for individual atoms is maximal, the Bragg signal has a local minimum because of absorption. This effect is similar to that seen in the experiment by Birkl et al.~\cite{birkl95}, except in that case the disorder was caused by temperature.

{\em AFM:}
The antiferromagnetic cloud is modeled by assigning lattice sites different pseudospin states according to N\'eel order.  We define $\Delta$ as the laser frequency detuning relative to the center of the two atomic resonances.
We assume a small Zeeman splitting $\Delta_0=\Gamma$ in the simulation so that the small sample of 925 atoms is optically dense when the probe laser is tuned directly between the resonances ($\mathrm{OD} = 7.4$).  This splitting is much smaller than the experimental parameters for $^6$Li near the Feshbach resonance, where $\Delta_0\approx 6.5 \Gamma $.  The optical density of the simulation is comparable to what one expects with $\sim 4\times 10^7$ $^6$Li atoms
\footnote{The optical density is estimated using the bulk properties of the gas: $\mathrm{OD} \approx ({2\omega}/{c})z \kappa_{\textrm{bulk}}(\Delta)$, where $z=N^{1/3}a$ is the thickness of the sample.}, about $40$ times the sample size expected in the experiment.

The results are shown for different radii $R$ (i.e.~different particle numbers) and disorder $\Delta x$ in Fig.~\ref{fig:multscatt}(e, f).  We find that the maximum intensity of the $(\half\;\half\;\half)$ Bragg peak scattered from an optically dense sample (e.g.~Fig.~\ref{fig:multscatt}(e)) is not found on one of the resonances (as one finds in the Born approximation or samples with low optical density), but rather around the midpoint in between the $\uparrow,\downarrow$ states ($\Delta \approx 0$).  Disorder increases incoherent scattering, reducing the amount of light coherently scattered into the Bragg peaks -- the disorder effectively enhances the optical density of the sample in a nonlinear way.

Under typical experimental conditions (finite temperature in a harmonic trap) the outer shell of the atomic cloud will be disordered and we expect the AFM order to develop only at the center. The disordered atoms could then shield the AFM from the probe light if the optical density in the shell becomes too high.  The above calculation demonstrates that in this case it is favorable to detune away from the $\uparrow,\downarrow$ states to increase the penetration depth of the light and maximize the signal at the $\left(\half\,\half\,\half \right)$ Bragg peak.
The physical picture is that for sufficiently large detuning (but between the two resonances) the disordered portion of the sample only contributes weak incoherent scattering, but the AFM portion still exhibits detectable coherent Bragg scattering from the magnetic planes.

\subsection{Symmetry effects}\label{sec:sym}
\begin{figure}
	\centering
		\includegraphics{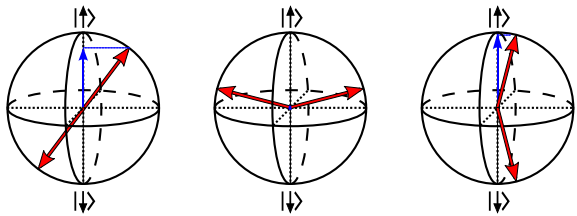}%
	\caption{\label{fig:bloch}(Color online) Cartoon of the Bloch sphere representations of the various AFM configurations.  Left: isotropic AFM state, where the spins (large red arrows) may lie anywhere on the Bloch sphere.  The $z$-component of the staggered spin vector is shown by the small blue arrow.  Center: canted AFM state, where the spins are restricted to be near the $xy$-plane but lift off of it by some small canting angle (exaggerated here for clarity).  In this case the $z$-projection of the staggered spin vector is zero.  Right: canted AFM state after applying a $\pi/2$-pulse.  Now the staggered spin has a nonzero $z$-component (small blue arrow), which may be detected by Bragg scattering.}
\end{figure}

The ordered state of a classical antiferromagnet is characterized by having a nonzero staggered magnetization $\mathbf{s}$.  In the absence of a symmetry breaking field, ${\mathbf{s}}$ will point in a random direction on the Bloch sphere (see Fig.~\ref{fig:bloch}), and sequential experiments will find this order pointing in a new direction.  Quantum mechanically the physics is similar, with the added feature that the staggered magnetization may be in a coherent superposition pointing in all directions.  From a practical standpoint, this distinction is minimal: a measurement will yield a definite (and random) direction for ${\mathbf{s}}$.
This (classical or quantum) uncertainty means that the 
the scattered intensity $I$ of a magnetic Bragg peak
will fluctuate from experiment to experiment.  The scattered light only interacts with the $\uparrow$ and $\downarrow$ eigenstates, and hence is only sensitive to the $z$-component of the order $s_z$.  Assuming an isotropically distributed $\mathbf{s}$, the 
 probability distribution of the intensity will be $f_{I}^{\textrm{iso}}(I)=\sqrt{I_{\mathrm{max}}/4I}$, with a mean value of $\bar{I}_{\textrm{iso}}=I_{\mathrm{max}}/3$, where $I_\textrm{max}$ is the intensity when the spins are parallel to $z$.  This purely geometric effect necessitates repeating the experiment multiple times.

Recent numerical calculations point out that a sufficient polarization of the sample ($|n_\uparrow-n_\downarrow|>0$) favors a staggered magnetization lying in the $xy$-plane of the Bloch sphere \cite{gottwald09,wunsch09}.  This response to a spin imbalance is analogous to the response of an electronic antiferromagnet to a magnetic field and can be explained in terms of a n anisotropic spin susceptibility.  As illustrated in Fig.~\ref{fig:bloch}, it is relatively easy to add a finite polarization to $xy$-antiferromagnetic order: one simply cants each spin slightly towards the $z$-axis of the Bloch sphere.  On the other hand, it is much more energetically costly to polarize a $z$-antiferromagnet: one must completely flip individual spins, adding $2J$ energy to the system for each spin that is flipped.  Thus, the lowest energy configuration of a polarized AFM has the N\'eel order in the $xy$-plane.  One expects that even at finite temperature the thermodynamic ensemble will be dominated by configurations with $\mathbf{s}$ lying in the $xy$-plane.  A simple model is to take the spins of the two sublattices to be
$\mathbf{s}_{1,2}=\frac{1}{2}\left[\pm(\cos\phi\cos\zeta)\hat{x}
\pm (\sin\phi\cos\zeta)\hat{y}
+ (\sin\zeta)\hat{z}\right]$, where $\phi$ is uniformly distributed over $2\pi$ radians of a circle and $\zeta\ll 1$ is a small \emph{fixed} canting angle relative to the $xy$-plane, which is set by the polarization.  

\begin{figure}
	\centering
		\includegraphics{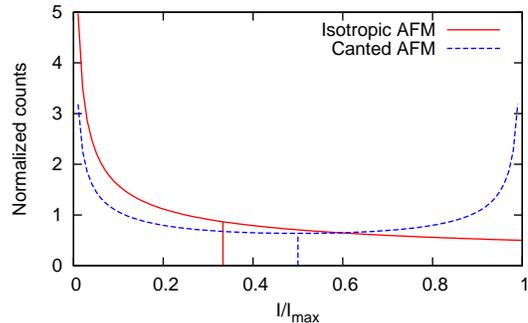}%
	\caption{\label{fig:Ipdf}(Color online)  Shot-to-shot probability distributions for the intensity of the Bragg scattering signal for two cases.  The solid red line shows the case when the staggered magnetization may point in any direction.  The dashed blue line shows the case for a canted AFM, after applying a $\pi/2$-pulse to rotate the spins from the $xy$-plane to the $xz$-plane.  The vertical lines indicate the mean values of the respective distributions.  The vertical scale is normalized such that the areas under both curves are one.}
\end{figure}

In the canted configuration, the $z$-component of the staggered magnetization is zero (center frame of Fig.~\ref{fig:bloch}), making the AFM invisible to Bragg scattering or optical imaging.  However, the spins may be reoriented to partially project along the $z$-axis by an RF ${\pi}/{2}$-pulse at the Zeeman splitting frequency, giving a non-zero Bragg signal (right panel of Fig.~\ref{fig:bloch}).  The resulting shot-to-shot intensity probability distribution of the Bragg scattered light is then  
$f_{I}^{\textrm{cant}}(I)=1/\sqrt{\pi^2(I/I_{\mathrm{max}})(1-I/I_{\mathrm{max}})}$
with a mean of $\bar{I}_{\textrm{cant}}=I_{\mathrm{max}}/2$.  The intensity distributions for Bragg scattering of the isotropic AFM and canted AFM (after applying the ${\pi}/{2}$-pulse) are shown in Fig.~\ref{fig:Ipdf}.  Note that the ${\pi}/{2}$-pulse does not change the probability distribution of the isotropic case because of symmetry.  Measuring these probability distributions provides a method to experimentally distinguish between the isotropic and canted AFM states, particularly because of the different  behavior of the two distributions when $I\sim I_\textrm{max}$.

\section{Conclusion}
These calculations confirm that optical Bragg diffraction is a viable method to observe antiferromagnetic ordering of atoms in an optical lattice.  The experimental measurement consists of using a CCD or photodiode to measure the intensities of the peaks.  The presence of the $(0\;0\;1)$ Bragg peak indicates cubic ordering, while the $(\half\;\half\;\half)$ Bragg peak indicates antiferromagnetic ordering.  In the actual experiment, only a portion of the atoms near the center of the crystal will be in the antiferromagnetic state. The ratio of the $(\half\;\half\;\half)$ intensity to the $(0\;0\;1)$ intensity provides a measure of the degree of ordering.   
Additionally, the shot-to-shot intensity distribution provides a method to determine whether the antiferromagnetic state is isotropic or canted.

\appendix*
\section{Full scattering formalism}\label{sec:scat}
Here we derive the full transition matrix (T-matrix), including multiple scattering events.  We consider scattering of a photon off of $N$ 2-level atoms located at positions $\mathbf{r}_j$ described by the matter-light Hamiltonian
\begin{equation*}
H= \underbrace{\sum_j \hbar \omega_{0,j} S_{z,j}+\sum_{\mathbf{k},\lambda} \hbar \omega_{\mathbf{k},\lambda} a^{\dagger}_{\mathbf{k},\lambda} a_{\mathbf{k},\lambda}}_{H_0}\underbrace{- \sum_j \mathbf{d}_j \cdot \mathbf{E}(\mathbf{r}_j)}_{V},
\end{equation*}
where $\hbar\omega_{0,j}$ is the energy splitting between the two levels, $S_{z,j}=\left( \ket{e}_j\bra{e}_j-\ket{g}_j\bra{g}_j \right)/2$ is the $z$-component spin operator for the $j$-th atom in terms of the corresponding ground state $\ket{g}_j$ and excited state $\ket{e}_j$, $\omega_{\mathbf{k},\lambda}$ is the frequency of a photon with wavevector $\mathbf{k}$ and polarization $\lambda$, $a^{\dagger}_{\mathbf{k},\lambda}$ and $a_{\mathbf{k},\lambda}$ are creation and annihilation operators of the same photon, $\mathbf{d}=e \mathbf{r}$ is the dipole operator ($e$ is the electron charge) and $\mathbf{E}(\mathbf{r})$ is the quantized electric field at the location of the atom at $\mathbf{r}$. In the rotating wave approximation (RWA) the matter-light interaction 
becomes \cite{gottfriedyan}
\begin{align*}
V=& -i \sum_{\mathbf{k},\lambda,j} g_{\mathbf{k}}
 \Big[ e^{i \mathbf{k} \mathbf{r}_j} (\mathbf{e}_{\mathbf{k},\lambda} \cdot \mathbf{e}^{*}_m) S_j^{+} a_{\mathbf{k},\lambda}\\
 &-e^{-i \mathbf{k} \mathbf{r}_j} (\mathbf{e}^{*}_{\mathbf{k},\lambda} \cdot \mathbf{e}_m) S_j^{-} a^{\dagger}_{\mathbf{k},\lambda}\Big],
\end{align*}
where $D \mathbf{e}_m^{*}=\bra{e}_j \mathbf{d}_j \ket{g}_j$, $\mathbf{e}_0=\hat{z}$, $\mathbf{e}_{\pm}=\mp (\hat{x} \pm i \hat{y})/\sqrt{2}$, $S_j^{+}=\ket{e}\bra{g}$, $S_j^{-}=\ket{g}\bra{e}$ and $g_{\mathbf{k}}=D \sqrt{2 \pi \hbar \omega_{\mathbf{k}}/L^3}$ ($L^3$ denotes the volume of the sample). In the low intensity limit (much less than the saturation intensity of the atomic transition) we may consider scattering of a single photon. Then within the RWA the Lippmann-Schwinger equation for the outgoing photon $\ket{\Psi^{(+)}}=\ket{{\mathbf{k}_i,\lambda_i}}+(E-H_0+i \epsilon)^{-1} V \ket{\Psi^{(+)}}$ (where $\ket{\mathbf{k}_i,\lambda_i}$ is an incoming plane wave photon with wavevector $\mathbf{k}_i$ and polarization $\lambda_i$, $\epsilon$ is a positive infinitesimal) closes with the ansatz
\begin{equation*}
\ket{\Psi^{(+)}}=\sum_{\mathbf{k},\lambda} \psi_{\mathbf{k},\lambda} \ket{\mathbf{k},\lambda}+\sum_j \psi_j S_j^{+}\ket{0},
\end{equation*}
and one obtains a linear equation for effective amplitudes $A_j$ (related to $\psi_j$ by $\psi_j=-i g_{k} \left( \mathbf{e}_{\mathbf{k}_j,\lambda_j}^{*} \cdot \mathbf{e}_m \right) A_j$)
\begin{equation}
\label{eq:lippmannschwinger}
\left( \Delta E_j+i \frac{\hbar \Gamma}{2} \right) A_j=e^{i \mathbf{k}_j \cdot \mathbf{r}_j}+\frac{\hbar \Gamma}{2} \sum_{l \neq j} \mathcal{G}_{jl} A_l,
\end{equation}
where $\Delta E_j$ is the (renormalized) detuning at site $j$ and $\Gamma=4 \omega_0^3 D^2/(3 \hbar c^3)$ is the linewidth of the transition. The dimensionless function $\mathcal{G}_{jl}=\beta_{jl}-i\gamma_{jl}$ has 
\cite{akkermans08} 
\begin{eqnarray*}
\beta_{jl}&=&\frac{3}{2} \left[-p \frac{\cos(k r_{jl})}{k r_{jl}}+q \left( \frac{\sin(k r_{jl})}{(k r_{jl})^2}+\frac{\cos(k r_{jl})}{(k r_{jl})^3} \right) \right],\\
\gamma_{jl}&=&\frac{3}{2} \left[p \frac{\sin(k r_{jl})}{k r_{jl}}+q \left( \frac{\cos(k r_{jl})}{(k r_{jl})^2}-\frac{\sin(k r_{jl})}{(k r_{jl})^3} \right) \right],
\end{eqnarray*}
where $r_{jl}=|\mathbf{r}_j-\mathbf{r}_l|$ and $k=|\mathbf{k}_j|=|\mathbf{k}_f|$ \footnote{Note that the RWA neglects retardation effects (or violates causality) and therefore our expression for $\mathcal{G}_{jl}$ is only valid when $k r_{jl} \gg 1$, which is the experimentally relevant limit.}, and $p$ and $q$ depend on whether one considers a $m=0$ or $m=\pm 1$ transition. One has $p=1-(\hat{z} \cdot \hat{\mathbf{r}}_{jl})^2,q=1-3 (\hat{z} \cdot \hat{\mathbf{r}}_{jl})^2$ for $m=0$ and $p=\frac{1}{2}(1+(\hat{z} \cdot \hat{\mathbf{r}}_{jl})^2),q=\frac{1}{2}(3(\hat{z} \cdot \hat{\mathbf{r}}_{jl})^2-1)$ for $m=\pm 1$ \cite{akkermans08}. The transition matrix is
\begin{eqnarray}
T_{\mathbf{k}_f,\lambda_f;\mathbf{k}_i,\lambda_i}&=&\big<{{\mathbf{k}_f,\lambda_f}}\big| V \big| {\Psi^{(+)}}\big>\notag\\
&=&g_k^2 \left( \mathbf{e}_{\mathbf{k}_f,\lambda_f} \cdot \mathbf{e}_m^{*}\right) \left( \mathbf{e}_m \cdot \mathbf{e}_{\mathbf{k}_i,\lambda_i}^{*}\right)\notag \\
& &\times\sum_j e^{-i \mathbf{k}_f \cdot \mathbf{r}_j}  A_j.\label{eq:fullT}
\end{eqnarray}
Eq.~\eqref{eq:lippmannschwinger} includes the effects of multiple scattering to all orders, provided one can solve for $A_j$. In the limit of low optical density of the sample, multiple scattering can be neglected and one obtains the Born approximation result $A_j=\left( \Delta E_j+i \frac{\hbar \Gamma}{2} \right)^{-1}e^{i \mathbf{k}_i \cdot \mathbf{r}_j}$ or
\begin{align*}
T_{\mathbf{k}_f,\lambda_f;\mathbf{k}_i,\lambda_i}=&g_k^2 \left( \mathbf{e}_{\mathbf{k}_f,\lambda_f} \cdot \mathbf{e}_m^{*}\right) \left( \mathbf{e}_m \cdot \mathbf{e}_{\mathbf{k}_i,\lambda_i}^{*} \right)\\
&\times \sum_j e^{i (\mathbf{k}_i-\mathbf{k}_f) \cdot \mathbf{r}_j} \left( \Delta E_j+i \frac{\hbar \Gamma}{2} \right)^{-1}.
\end{align*}
We finally note that the scattering amplitude $F_{\mathbf{k}_f,\lambda_f;\mathbf{k}_i,\lambda_i}$ is related to the T-Matrix through
\begin{eqnarray}
F_{\mathbf{k_f},\lambda_f;\mathbf{k_i},\lambda_i}&=&-\frac{L^3 k}{2 \pi \hbar c} T_{\mathbf{k_f},\lambda_f;\mathbf{k_i},\lambda_i}\notag \\
&=&-\frac{3}{2 k}\frac{\hbar \Gamma}{2} \left( \mathbf{e}_{\mathbf{k}_f,\lambda_f} \cdot \mathbf{e}_m^{*}\right) \left( \mathbf{e}_m \cdot \mathbf{e}_{\mathbf{k}_i,\lambda_i}^{*}\right)\notag \\
& &\times \sum_j e^{-i \mathbf{k}_f \cdot \mathbf{r}_j}  A_j\label{eq:fullF}.
\end{eqnarray}
In the Born approximation this reduces to Eq.~\eqref{eq:bornscatt}.

\begin{acknowledgments}
We thank David Huse, Han Pu, and Pedro Duarte for suggestions.  This work was supported under ARO Award W911NF-07-1-0464 with funds from the DARPA OLE Program.  Additional funding was provided by NSF, ONR, and the Welch and Keck Foundations. 
\end{acknowledgments}


\end{document}